%% file: main.tex
\documentclass[10pt, conference, letterpaper]{IEEEtran}

\input{header.tex}

\begin{document}
\include{cite}

\title{Shy Guys: A Light-Weight Approach to Detecting Robots on Websites}
\author{
  \IEEEauthorblockN{
    Rémi Van Boxem, Tom Barbette, Cristel Pelsser and Ramin Sadre
  }
  \IEEEauthorblockA{
    UCLouvain, \\
    Institute of Information and Communication Technologies, Electronics and Applied Mathematics, ICTEAM, \\
    Pôle en ingénierie informatique, INGI\\
    Place Sainte Barbe 2, Louvain-la-Neuve, Belgium\\
    Email: \{firstname.lastname\}@uclouvain.be
  }
}
\maketitle

\input{parts/abstract.tex}
\input{parts/introduction.tex}

\input{parts/related_work.tex}

\input{parts/methodology.tex}

\input{parts/experiments.tex}
\input{parts/conclusion.tex}
\input{parts/ethical_consideration.tex}

\printbibliography

\appendix

\input{parts/appendix.tex}

\end{document}

%% file: header.tex
\usepackage{url}
\usepackage{graphicx}
\usepackage{xcolor}
\usepackage{listings}
\usepackage{algorithm}
\usepackage{algpseudocode}
\usepackage{tikz}
\usepackage{subcaption}
\usepackage{pgfplots}
\pgfplotsset{compat=1.18}

\usepackage{booktabs}
\usepackage{siunitx}   %
\usepackage{bm}        %

\usepackage[style=ieee, backend=biber, maxbibnames=2, minbibnames=1]{biblatex}

\usepackage[english]{babel}
\usepackage{csquotes}
\usepackage{hyperref}
\usepackage{cleveref}
\AtEveryBibitem{%
  \clearfield{urlyear}\clearfield{urlmonth}\clearfield{urlday}%
  \clearfield{note}\clearfield{doi}\clearfield{isbn}\clearfield{issn} \clearfield{titleaddon}%
  \ifentrytype{article}{\clearfield{url}}{}%
  \ifentrytype{inproceedings}{\clearfield{url}}{}%
  \ifentrytype{report}{\clearfield{url}}{}%
}
\hypersetup{colorlinks=true, unicode=true, linkcolor=[rgb]{0.10,0.05,0.67}, citecolor=[rgb]{0.10,0.05,0.67}, filecolor=[rgb]{0.10,0.05,0.67}, urlcolor=[rgb]{0.10,0.05,0.67}}

\newcommand{\True}{\textbf{true}}
\newcommand{\False}{\textbf{false}}

\usetikzlibrary{shapes.geometric, arrows.meta, positioning, backgrounds, fit, calc}

\tikzset{
  block/.style={
    rectangle, rounded corners, minimum height=0.8cm,
    minimum width=1.4cm, text centered,
    draw=black, fill=white, font=\footnotesize,
  },
  decision/.style={
    diamond, aspect=2.8, text width=1.4cm,
    text centered, draw=black, fill=white,
    font=\footnotesize, inner sep=2pt,
  },
  arrow/.style={-{Stealth[length=2mm]}, thick},
  lbl/.style={font=\scriptsize, fill=white, inner sep=1pt},
}

\newcommand{\RVB}[1]{\textcolor{purple}{RVB: #1}}
\newcommand{\RS}[1]{\textcolor{blue}{RS: #1}}
\newcommand{\CP}[1]{\textcolor{olive}{CP: #1}}
\newcommand{\TB}[1]{\textcolor{orange}{TB: #1}}

\renewcommand{\RVB}[1]{}
\renewcommand{\RS}[1]{}
\renewcommand{\CP}[1]{}
\renewcommand{\TB}[1]{}

\newcommand{\etal}{\textit{et al.~}}

\usepackage{tikz}
\usepackage{xcolor}

\definecolor{cmTP}{RGB}{ 46, 134, 193}   %
\definecolor{cmFN}{RGB}{174, 214, 241}   %
\definecolor{cmFP}{RGB}{235, 245, 251}   %
\definecolor{cmTN}{RGB}{ 21,  67, 138}   %

\defbibfilter{academic}{
  type=article or
  type=inproceedings or
  type=incollection or
  type=thesis or
  type=report or
  type=book
}

\defbibfilter{other}{
  not type=article and
  not type=inproceedings and
  not type=incollection and
  not type=thesis and
  not type=report and
  not type=book
}

\definecolor{nightblue}{RGB}{0,33,78}
\definecolor{cyan}{RGB}{93,179,230}
\definecolor{grayblue}{RGB}{156,183,212}

\definecolor{cf1}{HTML}{E69F00}
\definecolor{cf2}{HTML}{56B4E9}

%% file: parts/abstract.tex
\begin{abstract}
Automated bots now account for roughly half of all web requests, and an increasing number deliberately spoof their identity to either evade detection or to not respect \texttt{robots.txt}. Existing countermeasures are either resource-intensive (JavaScript challenges, CAPTCHAs), cost-prohibitive (commercial solutions), or degrade the user experience. This paper proposes a lightweight, passive approach to bot detection that combines user-agent string analysis with favicon-based heuristics, operating entirely on standard web server logs with no client-side interaction. We evaluate the method on over 4.6 million requests containing 54,945 unique user-agent strings collected from websites hosted all around the earth. Our approach detects 67.7\% of bot traffic while maintaining a false-positive rate of 3\%, outperforming state of the art (less than 20\%). This method can serve as a first line of defense, routing only genuinely ambiguous requests to active challenges and preserving the experience of legitimate users.

\end{abstract}

%% file: parts/introduction.tex
\section{Introduction}

In late 2021, a theory called “Dead Internet Theory” began to gain traction on the internet. This theory suggests that a significant portion of internet traffic is generated by bots rather than humans \cite{tiffany_maybe_2021}. While often dismissed as a conspiracy theory, the latest annual report from Cloudflare indicates that bots (whether AI or non-AI) are responsible for half of requests to HTML pages, seven percentage points above human-generated traffic \cite{belson_2025_2025}. The same report also shows that traffic generated by AI crawlers increased to 15 times more than before.

In April 2025, Wikimedia, the organization behind Wikipedia, published a blog post explaining its struggle to keep up with bots \cite{mueller_how_2025}. While traffic spikes are not uncommon when major events occur and while Wikimedia previously handled such surges without difficulty, it explained that baseline traffic had increased by 50\% since January 2024 and that the organization has had greater difficulty serving content when major events (traffic spikes) occur.
Further analysis showed that 65\% of its most expensive traffic comes from bots.
Due to Wikimedia's infrastructure, expensive traffic is associated with less popular pages (less popular pages are not always cached and are requested from the main server). Scrapers tend to read larger numbers of pages in bulk than humans, not only the most popular ones, thereby defeating caching strategies.
One year later, the organization reported that it blocks or throttles approximately 25\% of all automated requests and yet continues to face challenges as a new generation of crawlers spoofs the identities of real web browsers and routes traffic through residential proxies to blend in with legitimate users \cite{mueller_quo_2026}.

This sudden increase in bot traffic can pose a serious threat to performance.
It is not uncommon to observe content creators struggling to protect their websites \cite{kwon2025web,Arlitt2001shopping}, forcing them to adopt CAPTCHA; migrate to hosting services that can detect bots and absorb the load of undetected bots; or employ aggressive measures that may block legitimate users (e.g., geo-blocking, mandatory JavaScript challenges) \cite{edwards_open_2025}. One of the most common methods to identify bots is to examine their self-reported user-agent string. However, this method is not very reliable, as bots can easily spoof their user-agent string to mimic that of a legitimate browser.

This paper proposes a lightweight, passive approach to bot detection that combines user-agent string analysis with favicon-based heuristics, operating entirely on standard web server logs with no client-side interaction. We evaluate our approach on various web server logs and show that it can serve as a first line of defense with a low false positive rate.

\Cref{sec:background-related} presents methods for detecting web crawlers and the background needed to understand the heuristics introduced in \cref{sec:methodology} and evaluated in \cref{sec:evaluation}. Finally, the results are discussed in \cref{sec:conclusion}, and ethical considerations related to log collection are addressed in \cref{sec:ethical-considerations}.

%% file: parts/related_work.tex
\section{Background and Related Work} \label{sec:background-related}

This section provides an introduction to the the user-agent header and a discussion of related work.

\subsection{Background}\label{sec:presua}

A user agent (UA) is defined as "the client which initiates a request" in RFC 2616 \cite{nielsen_hypertext_1999}. Usually, the user agent is a web browser, but it can also be a bot, a crawler, a mobile application, etc. A user-agent string is usually sent to the server via the "User-Agent" HTTP header \cite{fielding_http_2022, noauthor_user-agent_2025} and often contains information about the software and hardware of the client, such as the browser name and version, the operating system, the device type, etc.

\begin{lstlisting}[breaklines=true, basicstyle=\ttfamily\small, caption={[Mock User-Agent]User-Agent string sent by Safari on a MacBook Pro M5.}, label={lst:mock_user_agent}]
  Mozilla/5.0 (Macintosh; Intel Mac OS X 10_15_7) AppleWebKit/605.1.15 (KHTML, like Gecko) Version/26.3.1 Safari/605.1.15
\end{lstlisting}

\Cref{lst:mock_user_agent} shows a typical user-agent string that contains the following information: %

\begin{itemize}
  \item The first part \texttt{Mozilla/5.0} is a general token that says that the browser is Mozilla-compatible. For historical reasons, almost every browser today sends it. \cite{noauthor_user-agent_2025}
  \item The second part \texttt{(Macintosh; Intel Mac OS X 10\_15\_7)} contains information about the operating system and the device type. Here, it indicates that the client is running macOS 10.15.7 on an Intel-based Mac.
  \item The third part \texttt{AppleWebKit/605.1.15 (KHTML, like Gecko)} contains information about the rendering engine used by the browser. In this case, it indicates that the browser is using the WebKit rendering engine.
  \item The fourth part \texttt{Version/26.3.1 Safari/605.1.15} contains information about the browser name and version. In this case, it indicates that the browser is Safari version 26.3.1.
\end{itemize}

It is worth to note that the above example user-agent string was actually sent by a browser running on a MacBook Pro M5 and \emph{not} on an Intel-based Mac. Such a discrepancy can be due to one of two reasons: either the user-agent string is forged to avoid \texttt{robots.txt} restrictive rules \cite{kim_scrapers_2025}, or the user-agent reduction technique is used \cite{noauthor_user-agent-reduction_2025}.

\label{sec:ua-reduc}

User-Agent reduction is a technique used by some browsers to reduce the amount of information sent in the User-Agent header in order to protect users' privacy \cite{noauthor_user-agent-reduction_2025}. Usually, the user-agent is reduced to a generic string that does not contain sufficient information to fingerprint  clients. A common convention is to use fixed values for the platform version and the device model. The following values have been observed \cite{noauthor_user-agent-reduction_2025}:

\begin{itemize}
  \item \texttt{Android 10; K} on Android.
  \item \texttt{Macintosh; Intel Mac OS X 10\_15\_7} on macOS.
  \item \texttt{Windows NT 10.0; Win64; x64} on Windows.
  \item \texttt{X11; CrOS x86\_64 14541.0.0} on ChromeOS.
  \item \texttt{X11; Linux x86\_64} on Linux.
\end{itemize}

This initiative, led by Chromium, is part of an effort to reduce the amount of information sent by the user agent in order to protect the privacy of the users \cite{noauthor_user-agent_nodate,noauthor_user-agent_nodate-1,noauthor_wicgua-client-hints_2026}. Today, any modern browser \emph{should} send this generic information, and any user agent that does not follow this convention is likely to be a bot as demonstrated in \cref{sec:eval-ua}.

\RVB{Cite the result \cite{senol_unveiling_2023}}

\subsection{Detection of crawlers}

A \texttt{robots.txt} file is a text file placed on a web site that instructs web crawlers which resources they can or cannot request. Between 2022 and 2024, Liu \etal conducted a large-scale study of crawlers, analyzing \texttt{robots.txt} files on 40,000 websites and detecting that 14\% of them imposed restrictions on crawlers, particularly after OpenAI released its crawler in 2023 \cite{liu_somesite_2025}. However, as the authors point out, many crawlers do not respect \texttt{robots.txt} rules and therefore still access content even if they are undesired.
Kim \etal show that many crawlers tend to ignore strict rules in \texttt{robots.txt} and do not read it frequently, which means that even if a website owner adds a crawler in the file, the bot may continue to access the website content \cite{kim_scrapers_2025} for some time. Lastly, some crawlers can even go as far as to fake their user-agent string to mimic other crawlers, making it even harder to detect them \cite{kim_scrapers_2025}.

For now, the most common method to detect bots is by comparing their self-reported user-agent string to a list of known bots. Other methods include JavaScript challenges, CAPTCHAs, TLS fingerprinting (JA3/JA4), behavioural analysis, and commercial solutions:

\subsubsection{Active techniques}

To block bots, some websites use active techniques that require client-side JavaScript execution. For example, they can use JavaScript challenges to check whether the client can execute JavaScript code, which most bots cannot do. Anubis, well-known self-hosted option \cite{noauthor_techarohqanubis_2026}, presents a challenge under certain conditions and allows access to the content only if it is successfully completed. This challenge is based on Hashcash, a proof-of-work system that requires the client to perform a certain amount of computational work before being granted access \cite{back_hashcash_2002}.

Another approach is the use of CAPTCHAs (Completely Automated Public Turing test to tell Computers and Humans Apart).  CAPTCHAs are automated challenges designed to distinguish humans from bots \cite{goos_captcha_2003}, ranging from simple puzzles to behavioural analysis of user interactions \cite{shet_are_2024}.

Both approaches can be effective in preventing bots from accessing content, but they can also degrade the user experience for legitimate users. JavaScript challenges can cause delays in page loading and do not work for users with JavaScript disabled. CAPTCHAs can be frustrating for users, particularly when the problems are difficult to solve or present accessibility issues. Therefore, although active techniques can be effective in blocking bots, they should be used with caution to avoid negatively impacting the user experience.

\subsubsection{Passive header-based techniques}

TLS fingerprinting, such as JA3 and JA4, is a passive technique that analyzes the TLS handshake to identify clients based on their TLS fingerprints. The core idea is that when a client initiates an HTTPS connection, the initial handshake (specifically, the \texttt{ClientHello} message) is sent in plaintext and contains a list of supported cipher suites that varies across client applications \cite{husak_https_2016}.

After the TLS handshake, the client typically sends a user-agent string in the HTTP headers to identify the client application. Because many bots identify themselves as bots, crowdsourced lists exist to detect them. Examples include \cite{noauthor_arcjetwell-known-bots_2026, matomo_matomo-orgdevice-detector_2026, beech_jaybizzlecrawler-detect_2026} and \cite{noauthor_ai-robots-txtairobotstxt_2026}, which focus on bots and AI crawlers, with the latter directly providing block rules for NGINX, HAProxy, and Caddy. However, in its basic form, this method is not very reliable, as bots can spoof their user-agent.  %

Another technique is to check IP reputation lists to determine whether an IP address is well known for sending bots \cite{rohm_anthemakergoodbots_2026}.
Finally, behavioural techniques such as mouse-movement analysis \cite{wei_deep_2019} offer stronger guarantees by observing client interactions rather than self-declared headers, but require client-side instrumentation.

\subsubsection{Commercial solutions}

All the previously discussed techniques require development and installation on the server side and are more or less resource-intensive, depending on the technique employed. To avoid such work, turnkey solutions are available. These solutions, often outsourced and primarily provided by CDNs~\cite{noauthor_cloudflare_nodate,noauthor_bot_nodate}, are typically closed source and combine the previous techniques with others, such as intersite tracking based on browsing history patterns.

%% file: parts/methodology.tex
\section{Methodology} \label{sec:methodology}

\begin{figure*}[t]
  \centering
  \input{figures/methodology-overview.tex}
  \caption{Overview of the bot detection methodology. Raw logs from multiple web server formats are anonymized, normalized, and then analyzed through two complementary approaches: favicon-based analysis and user-agent header-based analysis.}
  \label{fig:methodology-overview}
\end{figure*}
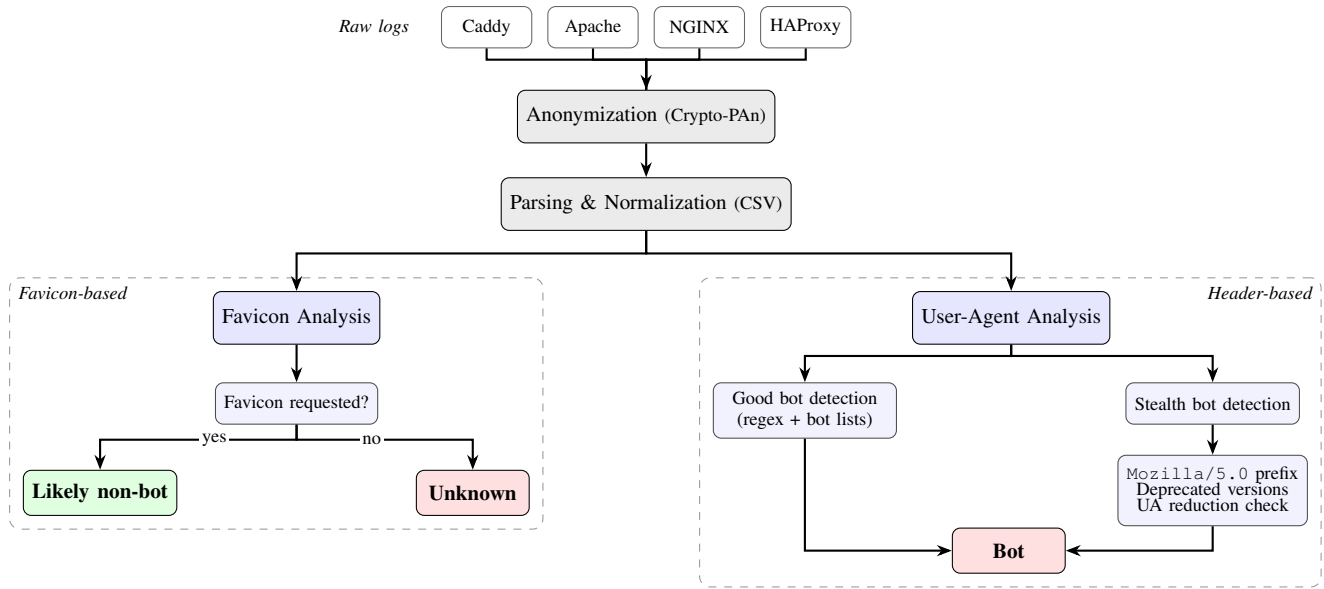

In this section, we describe our methodology to detect bots from web server logs. We propose two independent methods that we will then evaluate in \cref{sec:evaluation}. Both methods take as input information available in most web server log formats, such as the timestamp, client IP, HTTP method, URL, HTTP status code, and user-agent. We support Caddy, Apache HTTP Server (\emph{httpd}), NGINX, and HAProxy log formats. 
During our pre-processing phase, we first anonymize IP addresses using Crypto-PAn \cite{xu_prefix-preserving_2002}, then extract the fields of interest into CSV files. \Cref{fig:methodology-overview} shows the complete pipeline.

\subsection{Favicon} \label{sec:methfavicon}

A favicon is a small icon displayed in browser tabs and bookmarks to visually identify a website \cite{bamberg_favicon_2025}. Browsers cache it after the initial retrieval and rerequest it only after a defined interval or when the cache is cleared.

Our first detection method is based on how favicons are handled by human users and bots. The latter, especially those designed for web scraping, often use headless browsers that do not render the page visually. As a result, they do not request favicons at all. We therefore argue that the presence of a favicon request can serve as an indicator that the client is \emph{not} a bot.

However, false positives can occur with this method when clients cache the favicon and then later change their IP address and access the web server again without re-requesting the favicon. Such false positives can be reduced by forcing clients to regularly refetch the favicon. In the experiments in \cref{sec:eval-favicon}, we achieve this by daily changing the URL of the favicon on the investigated web server.

\subsection{User-agent} \label{sec:user-agent}

Our second method relies on an analysis of the user-agent string provided by the client in the HTTP requests. Because these strings are self-defined data, bot authors can choose whether to indicate that the client is a bot (by providing an explicit user-agent string) or to remain stealthy.

\subsubsection{Good bots} \label{sec:good-bots}

``Good'' bots identify themselves as bots in the user-agent string. For example, the bots of the Google search engine identify themselves as ``Googlebot'' \cite{noauthor_what_nodate}. We can assume that any user agent identifying itself as a bot is indeed a bot, unless it is forged and pretending to be someone else. To identify good bots, we search for words such as "bot", "crawler", or "spider" in the user-agent string using the regex patterns from the "\texttt{user-agents}" Python library \cite{ong_selwinpython-user-agents_2026}, hereafter referred to as "\texttt{py-ua}" Furthermore, we look the strings up in a list of known bots. This is necessary as some bots do not put the aforementioned words in their user-agent string. For example, one of OpenAI's bots identifies itself as \texttt{ChatGPT-User} \cite{noauthor_overview_nodate}. We rely on \texttt{ai-robots.txt}, a crowdsourced list that is updated regularly and focuses on AI related bots \cite{noauthor_ai-robots-txtairobotstxt_2026} (referred as "\texttt{robots.json}").
\CP{You'll have to explain somewhere in the paper why such lists exist. Here say why you use this list and not another one.}

\subsubsection{Stealthy bots} \label{sec:ua-coherence}

When a client does not identify itself as a bot in the user-agent string, it may be either a human user or a stealthy bot (aka: shy guy) designed to mimic a human user. Modern web browsers installed on end user devices such as PCs or smartphones follow certain rules in their user-agent strings that we exploit to discriminate humans from bots.

As cited in \cref{sec:presua}, user-agent strings from major web browsers \emph{should} start with the "Mozilla/5.0" prefix for legacy reasons \cite{andersen_history_2008}. User-agent strings deviating from this are unlikely to be sent by any modern web browser. For example, the default user-agent string for curl is \texttt{curl/VERSION} \cite{stenberg_curlcurl_1996}, although since version 4.5.1, curl allows setting a custom user-agent \cite{stenberg_curl_1998}. Similarly, monitoring solutions such as Nagios have a default user-agent string starting with \texttt{check\_http} \cite{noauthor_nagios-pluginsnagios-plugins_2026, noauthor_monitoring-pluginsmonitoring-plugins_2026}. We therefore treat any user-agent string that does not start with "Mozilla/5.0" as issued by a bot.

Another way to detect stealthy bots is to check for deprecated browser or operating system in the user-agent string. Wikimedia Foundation's user-agent breakdowns show that the percentage of users using deprecated browser versions is very low for both Google Chrome and Firefox \cite{wikimedia_foundation_user_2026}. Once a new version of a browser is released, the previous version is rapidly abandoned by users due to automated updates built in most browsers and operating systems. In 2023, all major browsers showed rapid adoption rates, with more than 80 \% of adoption peak in almost a week \cite{tijhof_browser_2023}. Therefore, we classify clients that report versions that have been deprecated for a long time (in our experiments: two years) as bots that either send fake user-agent strings or rely on old libraries or headless engines.

Finally, as stated in \cref{sec:ua-reduc}, any modern browser should send a reduced user-agent string that does not contain specific version information, so any user-agent string that contains such information is likely to originate from a bot.

Algorithm~\ref{alg:ua-detection} summarizes our detection method based on user-agent string analysis. It can be used alone or in combination with the favicon-based method.

\RS{Discuss forged user-agent string in discussion section}

\begin{algorithm}[t]
  \caption{User-Agent Based Bot Detection}
  \begin{algorithmic}[1]
    \Function{IsBot}{$ua$} \Comment{$ua$: user-agent string}
    \If{\Call{IsBotByRegex}{$ua$} \textbf{or} \Call{IsBotByList}{$ua$}}
    \State \Return \True
    \EndIf
    \If{$\neg\ ua$\Call{.startsWith}{"Mozilla/5.0"}}
    \State \Return \True
    \EndIf
    \If{\Call{HasDeprecatedVersion}{$ua$}}
    \State \Return \True
    \EndIf
    \State \Return \False
    \EndFunction
  \end{algorithmic}
  \label{alg:ua-detection}
\end{algorithm}

%% file: figures/methodology-overview.tex
\begin{tikzpicture}[
    node distance=0.5cm and 0.5cm,
    stage/.style={
      rectangle, rounded corners=3pt, minimum height=0.7cm,
      minimum width=2.2cm, text centered,
      draw=black, fill=black!8, font=\footnotesize,
    },
    source/.style={
      rectangle, rounded corners=3pt, minimum height=0.55cm,
      minimum width=1.2cm, text centered,
      draw=black!60, fill=white, font=\scriptsize,
    },
    analysis/.style={
      rectangle, rounded corners=3pt, minimum height=0.7cm,
      minimum width=2.2cm, text centered,
      draw=black, fill=blue!10, font=\footnotesize,
    },
    check/.style={
      rectangle, rounded corners=3pt, minimum height=0.55cm,
      minimum width=1.8cm, text centered,
      draw=black!70, fill=blue!5, font=\scriptsize,
    },
    result/.style={
      rectangle, rounded corners=3pt, minimum height=0.6cm,
      minimum width=1.5cm, text centered,
      draw=black, fill=green!12, font=\footnotesize\bfseries,
    },
    resultbot/.style={
      rectangle, rounded corners=3pt, minimum height=0.6cm,
      minimum width=1.5cm, text centered,
      draw=black, fill=red!12, font=\footnotesize\bfseries,
    },
    groupbox/.style={
      rectangle, rounded corners=4pt, draw=black!40,
      dashed, inner sep=5pt,
    },
    arrow/.style={-{Stealth[length=2mm]}, thick},
    lbl/.style={font=\scriptsize, fill=white, inner sep=1pt},
  ]

  \node [source]                        (caddy)   {Caddy};
  \node [source, right=0.2cm of caddy]  (apache)  {Apache};
  \node [source, right=0.2cm of apache] (nginx)   {NGINX};
  \node [source, right=0.2cm of nginx]  (haproxy) {HAProxy};

  \node [left=0.3cm of caddy, font=\scriptsize\itshape] (srclbl) {Raw logs};

  \coordinate (srcbot) at ($(apache.south)!0.5!(nginx.south)$);
  \node [stage, below=0.55cm of srcbot] (anon) {Anonymization {\scriptsize(Crypto-PAn)}};

  \draw [arrow] (caddy.south)   -- ++(0,-0.15) -| (anon.north);
  \draw [thick]  (apache.south)  -- ++(0,-0.15) -| (anon.north);
  \draw [thick]  (nginx.south)   -- ++(0,-0.15) -| (anon.north);
  \draw [thick]  (haproxy.south) -- ++(0,-0.15) -| (anon.north);

  \node [stage, below=0.45cm of anon] (parse) {Parsing \& Normalization {\scriptsize(CSV)}};
  \draw [arrow] (anon) -- (parse);

  \node [analysis, below left=0.8cm and 1.6cm of parse]  (favicon) {Favicon Analysis};
  \node [analysis, below right=0.8cm and 1.6cm of parse] (ua)      {User-Agent Analysis};

  \draw [arrow] (parse.south) -- ++(0,-0.3) -| (favicon.north);
  \draw [arrow] (parse.south) -- ++(0,-0.3) -| (ua.north);

  \node [check, below=0.5cm of favicon] (favreq) {Favicon requested?};
  \draw [arrow] (favicon) -- (favreq);

  \node [result, below left=0.6cm and 0.5cm of favreq]     (nonbot) {Likely non-bot};
  \node [resultbot, below right=0.6cm and 0.5cm of favreq] (unkfav) {Unknown};

  \draw [arrow] (favreq.south) -- ++(0,-0.2) -| node[lbl, near start, right] {yes} (nonbot.north);
  \draw [arrow] (favreq.south) -- ++(0,-0.2) -| node[lbl, near start, left]  {no}  (unkfav.north);

  \begin{pgfonlayer}{background}
    \node [groupbox, fit=(favicon)(favreq)(nonbot)(unkfav),
    label={[font=\scriptsize\itshape, anchor=north west]north west:Favicon-based}] (favgroup) {};
  \end{pgfonlayer}

  \node [check, text width=2.2cm, below left=0.5cm and 0.2cm of ua]  (goodbot)  {Good bot detection\\\scriptsize(regex + bot lists)};
  \node [check, below right=0.5cm and 0.2cm of ua] (stealth)  {Stealth bot detection};

  \draw [arrow] (ua.south) -- ++(0,-0.15) -| (goodbot.north);
  \draw [arrow] (ua.south) -- ++(0,-0.15) -| (stealth.north);

  \node [check, below=0.4cm of stealth] (stealthdetail) {
    \begin{tabular}{@{}c@{}}
      \scriptsize \texttt{Mozilla/5.0} prefix\\[-2pt]
      \scriptsize Deprecated versions\\[-2pt]
      \scriptsize UA reduction check
    \end{tabular}
  };
  \draw [arrow] (stealth) --  (stealthdetail);

  \coordinate (botmid) at ($(goodbot.south)!0.5!(stealthdetail.south)$);
  \node [resultbot, below=0.6cm of botmid] (bot) {Bot};

  \draw [arrow] (goodbot.south)       -- ++(0,-0.2) |- (bot.west);
  \draw [arrow] (stealthdetail.south) -- ++(0,-0.2) |- (bot.east);

  \begin{pgfonlayer}{background}
    \node [groupbox, fit=(ua)(goodbot)(stealth)(stealthdetail)(bot),
    label={[font=\scriptsize\itshape, anchor=north east]north east:Header-based}] (uagroup) {};
  \end{pgfonlayer}

\end{tikzpicture}

%% file: parts/experiments.tex
\section{Evaluation} \label{sec:evaluation}

We evaluate our proposed bot-detection approach across three dimensions. First, we assess the validity of the favicon as a proxy for authenticated traffic, verifying that it captures the same population of legitimate users without exposing any personally identifiable information. Second, we examine the complementary roles of regex-based rules and curated bot user-agent lists in identifying good bots, and quantify the coverage lost when either component is omitted. Third, we analyze the user-agent string signatures of bad bots whether these can reliably distinguish malicious crawlers from legitimate browsers. Lastly, we compare our resultant method with state-of-the-art methods.

To better study how bots behave across websites, two honeypots have been deployed on the internet, both accessible via domain name, IPv4 address, and IPv6 address. Both honeypots are static, one asks to be indexed by Google and the other is not indexed by any search engine. \texttt{robots.txt} files are configured to allow all bots to crawl the website for the first honeypot, while the second honeypot disallows all bots from crawling the website.

Other logs are collected from third-party partners, including a Learning Management System (LMS). Our logs are provided from websites hosted in different locations (Japan, USA, Europe,\dots). These logs are collected directly from the web server and contain only HTTP headers and metadata about the requests; they are not enriched with other information.

The dataset used for evaluation consists of web traffic logs collected from the aforementioned sources over a period of several years, starting in July 2024. The logs contain a total of 4,594,072 requests, with 54,945 unique user-agent strings. The dataset is anonymized to protect the privacy of individuals and organizations involved in the study, following the ethical considerations outlined in \cref{sec:ethical-considerations}.

\subsection{Favicon}\label{sec:eval-favicon}

The ground truth is composed of an estimation of the number of \emph{anonymized} authenticated IPs for the LMS logs. To avoid complex fingerprinting matching IPs to a specific user, which can increase the risk of privacy breaches and might not respect the ethical considerations outlined in \cref{sec:ethical-considerations}, another, less precise approach is used.

By design, most content on the LMS is accessible only to authenticated users. A user can read and view some materials without registering, but must be registered and logged in to submit and test their work. Students submit their work to the LMS for grading via the web interface. Under the hood, the browser sends a POST request to a specific endpoint (e.g., \texttt{/course/}) to submit the work. An unauthenticated user \emph{post}ing to this endpoint receives a 404 status code, whereas an authenticated user receives a 200 status code. Counting the number of unique IP addresses that issue POST requests to the \texttt{/course/} endpoint provides an estimate of the number of unique authenticated users. This estimate is not perfect, as some users might share the same IP address (e.g., behind a NAT\footnote{The university from which the dataset comes does not use any kind of NAT.}), some users can access the LMS without submitting any code\footnote{Due to how this LMS is used, this behaviour is uncommon.}, and some users might have dynamic IP addresses. This number still provides a baseline to evaluate the performance of our detection method while respecting the ethical considerations outlined in \cref{sec:ethical-considerations}.

\begin{figure}[t]
  \centering
  \input{figures/favicon-post-ips.tex}
  \caption{Daily unique IP addresses issuing favicon requests and POST requests to \texttt{/course/} over the observation period.}
  \label{fig:daily_ips}
\end{figure}
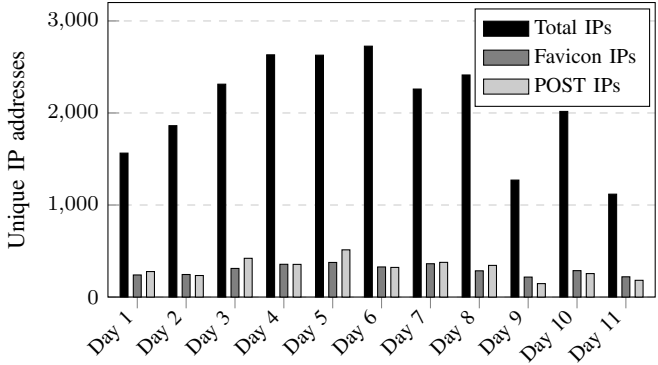

As stated in \cref{sec:methfavicon}, a CRON task ran daily to update the favicon URL (\texttt{/favicon.ico?v=DATE}), forcing browsers to request it daily. \Cref{fig:daily_ips} shows the number of unique IP addresses that requested the favicon, issued POST requests to the \texttt{/course/} endpoint, and the total number of unique IP addresses that issued any kind of request to the LMS on a daily basis. The numbers of unique IP addresses are similar for both the favicon requests and the POST requests, whereas the total number of unique IP addresses is much higher. The difference between the number of unique IP addresses that requested the favicon and the number that issued POST requests to the \texttt{/course/} endpoint can be explained by the fact that some authenticated users might not have submitted any work during the observation period or requested the favicon from another IP address (e.g., mobile network, home network, etc.). However, the similarity between the number of unique IP addresses that requested the favicon and the number that issued POST requests to the \texttt{/course/} endpoint suggests that the detection method for bots based on favicon requests is effective.

A paired t-test found no significant difference between the two daily counts ($t(10) = -0.981, p = 0.350$, Cohen's $d = -0.3$), indicating a small and non-significant effect. The negative $t$ value indicates that, on average, the first variable (favicon IPs) was slightly lower than the second (POST IPs), which corroborates the fact that the favicon is retrieved only once per browser per day, whereas the user may have changed IPs multiple times during the day and issued multiple POST requests. Furthermore, a Pearson correlation revealed a strong positive association between the two series ($r = 0.87, p < 0.001$), suggesting that the number of unique IP addresses requesting the favicon is strongly correlated with the number of unique IP addresses issuing POST requests to the \texttt{/course/} endpoint.

A solution to improve the accuracy of this detection method, given that our university network causes extensive IP hopping, would be to use fingerprinting techniques to match the IP addresses that requested the favicon with the IP addresses that issued POST requests to the \texttt{/course/} endpoint (session cookies, TLS fingerprints, \dots). However, this approach may increase the risk of privacy breaches and is thus incompatible with our ethical stance.

\subsection{User-agent} \label{sec:eval-ua}

\begin{table}[t]
  \caption{Top 10 most frequent user agents in the logs}
  \label{tab:top10ua}
  \centering
  \setlength{\tabcolsep}{5pt}
\begin{tabular}{@{}l l
  S[table-format=8,group-separator={,}]
  @{\,(}
  S[table-format=3.2]
  @{\,\%)}@{}}
  \toprule
  \textbf{Browser} & \textbf{Operating System}
    & \multicolumn{2}{c@{}}{\textbf{Requests}} \\
  \midrule
  Google Chrome 139          & Windows 10  & 1699193 & 36.98 \\
  Apache-CXF 3.5.8           & {---}       &  236298 &  5.14 \\
  Facebook External Agent & {---}       &  176824 &  3.84 \\
  Amazonbot 0.1              & {---}       &  144945 &  3.15 \\
  Google Chrome 142          & Windows 10  &  142854 &  3.10 \\
  Bingbot 2.0                & {---}       &  121909 &  2.65 \\
  ClaudeBot 1.0              & {---}       &  108189 &  2.35 \\
  Internet Explorer 8.0      & Windows XP  &   93610 &  2.03 \\
  {(empty)}                  & {---}       &   89875 &  1.95 \\
  Google Chrome 142          & macOS       &   80994 &  1.76 \\
  Other                      & {---}       &1699381 & 36.99 \\
  \midrule
  Total & {---} &4594072 & 100 \\
  \bottomrule
\end{tabular}

  \smallskip
  \raggedright
  \footnotesize
\end{table}

\Cref{tab:top10ua} lists the top 10 most frequent user agents in the logs. This top 10 includes well-known bots such as Googlebot, Bingbot, Amazonbot, and ClaudeBot, as well as a large number of requests with an empty user-agent string ("-"). Those user-agents account for a significant portion of the traffic representing approximately 63\% of the total requests. 

We validate the different techniques based on the user-agent string separately for good bots and stealthy robots:

\subsubsection{Good bots} \label{sec:eval-goodbots}

As explained in \cref{sec:good-bots}, good bots identify themselves as bots in the user-agent header. Cloudflare, one of the largest security and CDN providers, publishes a comprehensive directory of "certified bots" \cite{noauthor_verified_2025,noauthor_bots_nodate}. However, only a fraction of bots are listed in this directory \cite{noauthor_bots_2026}; the others are kept internal. This directory is used as a baseline to evaluate the true positivity of our detection method for good bots. The results of our detection method are compared with the list of bots provided by Cloudflare.

Among 54,945 unique user-agents, 373 are identified as robots by using the \texttt{py-ua} and \texttt{robots.json} techniques presented in \cref{sec:good-bots}. \Cref{tab:good-bots-upsetplot} shows the distribution and overlap among the different detection methods for good bots. Of these 373 user-agents, 185 were also identified as bots by Cloudflare, while 188 were not listed in the Cloudflare directory. Forty-one Cloudflare-listed bots were not detected by our method, yielding a recall of approximately $81.7\%$ with respect to Cloudflare as the ground truth.

Although 373 user-agents identified as bots by our method seems low (0.68\% of the total unique user-agents), weighting these user-agents by their frequency in the logs shows that they represent a significant portion of the traffic. Specifically, these 373 user-agents account for approximately $25.6\%$ of the total requests (with less than $0.2\%$ undetected bots) made on all captured traffic.  \Cref{fig:weighted-bots-upsetplot} in appendix shows the distribution by reason of flagged user agent weighted by frequency.

\begin{table}[t]
    \centering
    \caption{Overlap between our bot detection methods (\texttt{robots.json} and \texttt{py-ua} defined in \cref{sec:good-bots}) and Cloudflare directory across 373/414 detected bot user-agents, out of 54,945 unique user-agents.}
    \label{tab:good-bots-upsetplot}
    \setlength{\tabcolsep}{5pt}
    \begin{tabular}{c c c r}
    \toprule
    \textbf{\texttt{robots.json}} \cite{noauthor_ai-robots-txtairobotstxt_2026} & \textbf{\texttt{py-ua}} \cite{ong_selwinpython-user-agents_2026} & \textbf{Cloudflare} \cite{noauthor_bots_nodate} & \textbf{Count} \\
    \midrule
        $\bullet$ & $\cdot$& $\cdot$& 111\\
         $\cdot$& $\bullet$& $\bullet$& 98 \\
         $\cdot$&$\bullet$&&90\\
         $\cdot$&$\cdot$&$\bullet$&41\\
         $\bullet$&$\bullet$&&28\\
         $\bullet$&$\bullet$&$\bullet$&25\\
         $\bullet$&$\cdot$&$\bullet$&21\\
         \midrule
         \multicolumn{1}{l}{\textit{Set sizes:}  185} \hfill& 241 & 185 & 414 \\
    \bottomrule 
    \end{tabular}
\end{table}

Looking at the self-declared user-agent strings is not sufficient to detect all bots, as some bots do not identify themselves as such in the user-agent string. Such bots have usually "badly-forged" user-agents, which can be detected using the user-agent coherence method described in \cref{sec:ua-coherence}. The next section evaluates the performance of this method for detecting stealthy robots.
\TB{I am a bit lost here. Figure 3 only relates to the user agent coherence?. The 0.2pc of undecteded bots are known how? I wonder if we should not jump directly to figure 4?} \RVB{I am not sure to have fully understood the problem.}

\subsubsection{Stealth robots} \label{sec:eval-hiddenbot}

\Cref{fig:unweighted-bots-upsetplot} shows the distribution and overlap among the different detection methods for stealthy robots using an UpSet plot \cite{lex_upset_2014} for a unique set of user-agents. Among 54,945 unique user-agents, 51,268 (93\%) are identified as bots by the user-agent coherence method. Of these 51,268 user-agents, 48,646 (94.9\%) are identified as bots by the "deprecated browser" method, 27,496 (53.6\%) are identified as bots by the "deprecated OS" method, 1,688 (3.3\%) do not start with "Mozilla/5.0", 719 (1.4\%) are identified as bots by the "\texttt{user-agents}" library, and 241 (0.47\%) are found in the "\texttt{robots.json}" list.

The "deprecated" heuristic methods (browser and OS) might look like the more "aggressive" ones as they identify a large number of user-agents as bots. One might think they may generate numerous false positives. However, the majority of flagged user-agents by these methods are using more than 5 years old browsers and operating systems, which are unlikely to be used by legitimate users \cite{tijhof_browser_2023}. In these cases, it is important to pair our detection method with a more precise active one to avoid blocking someone arbitrarily.

\begin{figure}[t]
  \centering
  \input{figures/uau_android_versions.tex}
  \caption{Distribution of claimed Android versions in user-agent strings.}
  \label{fig:ua-android-versions}
\end{figure}
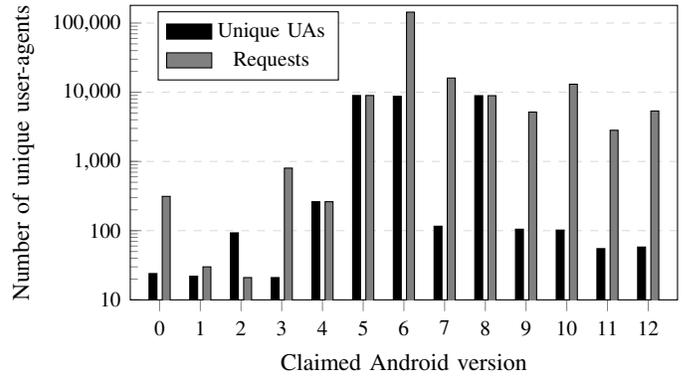

\Cref{fig:ua-android-versions} shows the distribution of claimed Android versions in user-agent strings. While a spike should be expected for Android 10 (released in September 2019) because it is the default platform information for Android devices used in user-agent reduction, the distribution shows that many user-agents claim to use Android versions that are no longer supported by Google (e.g., Android 4.4, released in 2013). According to the Android version distribution, Android 10 and later have a 90.8\% cumulative share \cite{belinski_android_nodate}, which is not reflected in the user-agent strings, suggesting that many of them use outdated operating systems or forged user-agent strings.

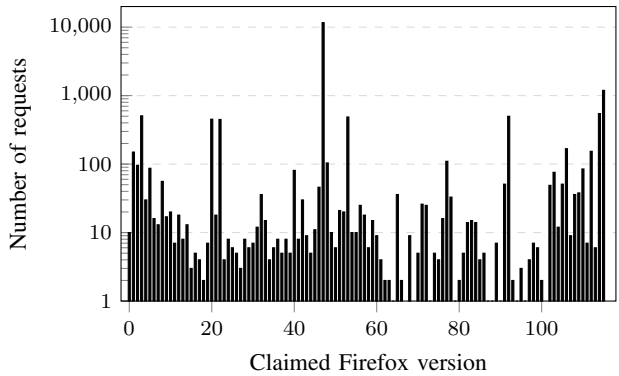
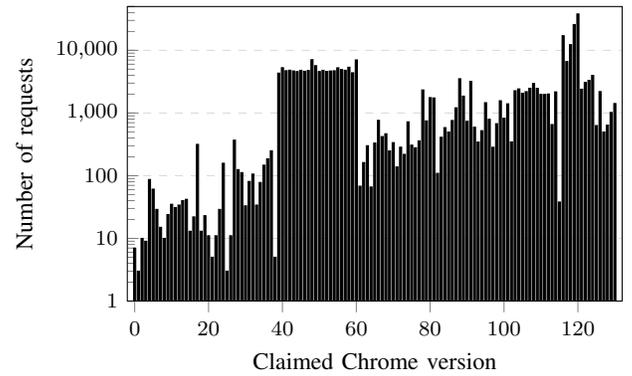
\begin{figure*}[htbp]
  \centering
  \begin{subfigure}{0.45\textwidth}
    \centering
    \input{figures/firefox_version.tex}
    \caption{Distribution of claimed Firefox versions in user-agent strings.}
    \label{fig:ua-firefox-versions}
  \end{subfigure}%
  \hfill
  \begin{subfigure}{0.45\textwidth}
    \centering
    \input{figures/chrome_version.tex}
    \caption{Distribution of claimed Chrome versions in user-agent strings. Chrome and Chrome Mobile are merged for readibility.}
    \label{fig:ua-chrome-versions}
  \end{subfigure}
  \caption{Distribution of claimed Chrome and Firefox versions in user-agent strings.}
\end{figure*}

Figures \ref{fig:ua-firefox-versions} and \ref{fig:ua-chrome-versions} show similar distributions for Firefox and Chrome versions, respectively, with many user-agents claiming to use outdated versions of these browsers.

For Firefox, a spike can be observed for version 47, which was released in June 2016 and ceased to receive updates in September 2017. Inside User-agents claiming to use Firefox 47, the first most common user-agent "\texttt{Mozilla/5.0 (Windows NT 6.1; Win64; x64; rv:47.0) Gecko/20100101 Firefox/47.0}" accounts for 11,401 requests and is the one given as an example by Mozilla in their documentation for user-agent parsing \cite{noauthor_user-agent_2025}. It is also available as an example user-agent used by Colly, a scraping library \cite{ascimoo_most_2018}.

For Chrome versions, a pronounced plateau can be observed for both desktops and mobiles around versions 39 and 60. Chrome 39, the first version of Google Chrome supporting 64-bit architectures, was released in November 2014 and ceased to receive updates in January 2015 \cite{google_stable_nodate}. Chrome 61 released in September 2017 was the first version to support JavaScript Modules natively \cite{lepage_new_2017}.

Similarly, our logs contain 558 unique user-agents claiming to use Internet Explorer (MSIE), officially discontinued by Microsoft in June 2022 and running versions that did not support TLS. The user-agent string \texttt{Mozilla/4.0 (compatible; MSIE 8.0; Windows NT 5.1; Trident/4.0)} is an Internet Explorer 8.0 user-agent, released in March 2009. This exact user-agent accounts for 93610 requests (2\% of the total) and is the eighth most prevalent user-agent in our logs. These requests are highly unlikely to originate from legitimate users. Further investigation indicates that web forums identify this user-agent as used by Nessus \cite{f5_web_nodate, noauthor_http_nodate}, a widely used vulnerability scanner.

We previously showed in \cref{sec:eval-favicon} that the favicon can be used as a reliable heuristic without compromising user privacy: the number of IPs requesting the favicon per day and the number of authenticated IPs agree in both trend and level across days. We also showed in \cref{sec:eval-goodbots} that both regex-based rules and a list of bot user-agent strings are required: the former prevents missing bots that do not include an obvious keyword in their self-declared string, and the latter catches bots that are not present in our list (for example, if the list is not updated). Lastly, we showed in \cref{sec:eval-hiddenbot} that bad bots usually use malformed user-agent strings or deprecated browsers that are unlikely to function on modern websites.

\subsection{Comparison against state-of-the-art}

To validate our results, and because no single fully labelled dataset contains both humans and bots, we constructed one by combining two reliable sources. Human traffic was drawn from the LMS, where real users were identified using the technique described in \cref{sec:eval-favicon} (augmented with internal traffic and filtered for monitoring requests). Bot traffic was drawn from the honeypot, which real users have no reason to access. The honeypot was considered a reliable source of bot traffic: the five most frequent origin providers, Microsoft, Cloudflare, Amazon, DigitalOcean, and Google Cloud Platform, accounted for 71\% of its requests. This combined dataset serves as ground truth for evaluating whether our heuristics correctly distinguish humans from bots.

The confusion matrix (\Cref{tab:comparison}) reveals a favourable asymmetry: the false positive rate is low at 3.0\%, meaning that only a small fraction of legitimate human requests are incorrectly flagged. The false negative rate is higher at 32.4\%, indicating that a nontrivial share of bot requests evades detection. This is an acceptable trade-off in a context where flagging a real user is more disruptive than missing a bot. That's why it is important to pair our method with a second line of defense.

Importantly, this recall of 67.6\% substantially outperforms a naive string-matching approach. For reference, the Cloudflare bot pattern list identifies only 8.4\% of requests in our log, covering 58 of the 873 unique user agents observed (6.6\%). By contrast, our method flags 491 unique user agents (56.2\%), an eightfold increase in coverage. This suggests that a large proportion of crawlers active in the infrastructure deliberately omit or spoof their user-agent strings to avoid rule-based block lists, and that a suspicion-based approach is necessary to surface them. It also significantly outperforms CrawlerDetect \cite{beech_jaybizzlecrawler-detect_2026} (18.1\%), Matomo DeviceDetector \cite{matomo_matomo-orgdevice-detector_2026} (12.0\%), and known-bot IP lists \cite{rohm_anthemakergoodbots_2026} (18.0\%), all of which miss over 80\% of bot requests in our dataset.

Well-behaved bots are expected to request the \texttt{robots.txt} file to check whether they are allowed to crawl a website~\cite{kim_scrapers_2025,koster_robots_2022}. However, in the "Bots" dataset, only 351 of 3,185 unique IPs (11\%) requested it, suggesting that the vast majority deliberately avoid identifying themselves as crawlers. Similarly, only 264 (8\%) requested the favicon, reinforcing the idea that most bots do not behave like real browsers and that the favicon heuristic may be a reliable discriminator.

While our method does not achieve perfect separation (a 32.4\% false negative rate indicates that some bots still pass undetected), the primary objective is not to intercept every automated request but to reduce the volume of traffic that must be subjected to active verification challenges. By confidently clearing 97.0\% of human requests and flagging 67.6\% of bots at the filtering stage, the method substantially narrows the population that needs further scrutiny. Consequently, the friction imposed by challenge mechanisms is reserved to a much smaller and genuinely ambiguous subset of requests, preserving the experience of the vast majority of legitimate users.

\begin{table}[t]
  \caption{Comparison of Bot Detection Methods}
  \label{tab:comparison}
  \centering
  \setlength{\tabcolsep}{5pt}
  \begin{tabular}{@{}l
      S[table-format=2.1]
      S[table-format=2.1]
      S[table-format=1.1]
      S[table-format=3.1]@{}}
    \toprule
    \textbf{Method}
      & {\textbf{TP\,(\%)}}
      & {\textbf{FN\,(\%)}}
      & {\textbf{FP\,(\%)}}
      & {\textbf{TN\,(\%)}} \\
    \midrule
    \textit{Our method} & \bfseries \textbf{67.6} & \bfseries \textbf{32.4} & \bfseries 3.0 & \bfseries 97.0 \\
    \midrule
    Cloudflare Bot Management~\cite{noauthor_bots_nodate}   &  8.4 & 91.6 & 0.1 & 99.9 \\
    CrawlerDetect~\cite{beech_jaybizzlecrawler-detect_2026}                & 18.1 & 81.9 & 0.1 & 99.9 \\
    Matomo DeviceDetector~\cite{matomo_matomo-orgdevice-detector_2026}            & 12.0 & 88.0 & 0.1 & 99.9 \\
    Known-bots IP list~\cite{rohm_anthemakergoodbots_2026}            & 18.0 & 82.0 & \textbf{0.0} & \textbf{100.0} \\
    \bottomrule
  \end{tabular}

  \smallskip
  \raggedright
  \footnotesize
  Percentages are row-normalised within each class (Bot / Human).
  TP and FN are computed over the Bot class; FP and TN over the Human class.
  The separator line distinguishes our method from state-of-the-art baselines.
\end{table}

\TB{It's unclear to me what is definitely for sure a bot. What might be a false detection? What should be checked manually to verify our method?
  What is the final proposed scheme to classify?
  If I take 58K requests.
How many are detected as bots by favicon and user-agent? Of those, how much do we miss using other techniques? Then, how much might be true positive?}

\TB{Maybe we should propose variants according to the mitigation in place. Likely bots -> go to Anubis etc.
Surely bots for blocking -> no false positive.?}

%% file: figures/favicon-post-ips.tex
\begin{tikzpicture}
  \begin{axis}[
      width=\columnwidth,
      height=5.5cm,
      ybar,
      area legend,
      bar width=3pt,
      ylabel={Unique IP addresses},
      symbolic x coords={Day 1,Day 2,Day 3,Day 4,Day 5,Day 6,Day 7,Day 8,Day 9,Day 10,Day 11},
      xtick=data,
      xticklabel style={rotate=45, anchor=east, font=\footnotesize},
      yticklabel style={font=\footnotesize},
      ylabel style={font=\small},
      xtick pos=bottom,
      ytick pos=left,
      xlabel style={font=\small},
      ymin=0,
      ymax=3200,
      ymajorgrids=true,
      grid style={dashed, gray!40},
      legend style={
        at={(0.67,0.98)},
        anchor=north west,
        font=\footnotesize,
        draw=black,
        fill=white,
        legend columns=1,
      },
      legend cell align={left},
      enlarge x limits=0.06,
    ]

    \addplot[
      fill=black,
      draw=black,
    ] coordinates {
      (Day 1,1564)
      (Day 2,1862)
      (Day 3,2313)
      (Day 4,2632)
      (Day 5,2628)
      (Day 6,2726)
      (Day 7,2260)
      (Day 8,2413)
      (Day 9,1271)
      (Day 10,2017)
      (Day 11,1118)
    };

    \addplot[
      fill=black!50,
      draw=black,
    ] coordinates {
      (Day 1,240)
      (Day 2,245)
      (Day 3,311)
      (Day 4,356)
      (Day 5,376)
      (Day 6,327)
      (Day 7,362)
      (Day 8,285)
      (Day 9,217)
      (Day 10,287)
      (Day 11,220)
    };

    \addplot[
      fill=black!20,
      draw=black,
    ] coordinates {
      (Day 1,277)
      (Day 2,234)
      (Day 3,421)
      (Day 4,355)
      (Day 5,513)
      (Day 6,322)
      (Day 7,377)
      (Day 8,344)
      (Day 9,146)
      (Day 10,255)
      (Day 11,182)
    };

    \legend{Total IPs, Favicon IPs, POST IPs}
  \end{axis}
\end{tikzpicture}

%% file: figures/uau_android_versions.tex
\begin{tikzpicture}
  \begin{axis}[
      width=\columnwidth,
      height=5.5cm,
      ybar,
      area legend,
      bar width=3pt,
      ymode=log,
      log ticks with fixed point,
      ylabel={Number of unique user-agents},
      xlabel={Claimed Android version},
      xtick=data,
      xticklabels={0,1,2,3,4,5,6,7,8,9,10,11,12},
      xticklabel style={font=\footnotesize},
      yticklabel style={font=\footnotesize},
      ylabel style={font=\small},
      xlabel style={font=\small},
      xtick pos=bottom,
      ytick pos=left,
      ymin=10,
      ymax=180000,
      ytick={10,100,1000,10000,100000},
      yticklabels={10,100,{1,000},{10,000},{100,000}},
      nodes near coords,
      nodes near coords align={vertical},
      point meta=explicit symbolic,
      enlarge x limits=0.06,
      grid=major,
      grid style={dashed, gray!30},
      major grid style={line width=0.2pt},
      ymajorgrids=true,
      xmajorgrids=false,
      legend style={
        at={(0.05,0.98)},
        anchor=north west,
        font=\footnotesize,
        draw=black,
        fill=white,
        legend columns=1,
      },
    ]
    \addplot[
      fill=black,
      draw=black,
    ] coordinates {
      (0,  24)   %
      (1,  22)   %
      (2,  93)   %
      (3,  21)   %
      (4,  262)  %
      (5,  8972) %
      (6,  8738) %
      (7,  116)  %
      (8,  8928) %
      (9,  105)  %
      (10, 102)  %
      (11, 55)   %
      (12, 58)   %
    };

    \addplot[
      fill=black!50,
      draw=black,
    ] coordinates {
      (0,  313)   %
      (1,  30)   %
      (2,  21)   %
      (3,  801)   %
      (4,  262)  %
      (5,  8972) %
      (6,  143686) %
      (7,  16050)  %
      (8,  8928) %
      (9,  5174)  %
      (10, 13064)  %
      (11, 2823)   %
      (12, 5361)   %
    };

    \legend{Unique UAs, Requests}
  \end{axis}
\end{tikzpicture}

%% file: figures/firefox_version.tex
\begin{tikzpicture}
  \begin{axis}[
      width=\columnwidth,
      height=5.5cm,
      ybar,
      bar width=0.8pt,
      ymode=log,
      log ticks with fixed point,
      ylabel={Number of requests},
      xlabel={Claimed Firefox version},
      xticklabel style={font=\footnotesize},
      yticklabel style={font=\footnotesize},
      ylabel style={font=\small},
      xlabel style={font=\small},
      xmin=-2, xmax=118,
      ymin=1,
      ymax=20000,
      xtick pos=bottom,
      ytick pos=left,
      grid=major,
      grid style={dashed, gray!30},
      major grid style={line width=0.2pt},
      ymajorgrids=true,
      xmajorgrids=false,
    ]
    \addplot[fill=black, draw=black] coordinates {
      (0,10) (1,150) (2,96) (3,507) (4,30) (5,87) (6,16) (7,13) (8,56) (9,17) (10,20) (11,7) (12,18) (13,8) (14,13) (15,3) (16,5) (17,4) (18,2) (19,7) (20,453) (21,18) (22,449) (23,4) (24,8) (25,6) (26,5) (27,3) (28,8) (29,6) (30,7) (31,12) (32,36) (33,15) (34,4) (35,6) (36,8) (37,5) (38,8) (39,5) (40,81) (41,8) (42,30) (43,9) (44,5) (45,11) (46,46) (47,11637) (48,104) (49,10) (50,6) (51,21) (52,20) (53,488) (54,10) (55,10) (56,25) (57,18) (58,6) (59,15) (60,9) (61,4) (62,2) (63,2) (64,1) (65,36) (66,2) (68,9) (70,5) (71,26) (72,25) (73,1) (74,5) (75,4) (76,16) (77,110) (78,33) (79,1) (80,2) (81,5) (82,14) (83,15) (84,14) (85,4) (86,5) (87,1) (88,1) (89,7) (90,1) (91,51) (92,500) (93,2) (94,1) (95,3) (96,1) (97,4) (98,7) (99,6) (100,2) (102,49) (103,76) (104,12) (105,51) (106,168) (107,9) (108,36) (109,38) (110,85) (111,7) (112,154) (113,6) (114,548) (115,1194)
    };
  \end{axis}
\end{tikzpicture}

%% file: figures/chrome_version.tex
\begin{tikzpicture}
  \begin{axis}[
      width=\columnwidth,
      height=5.5cm,
      ybar,
      bar width=0.8pt,
      ymode=log,
      log ticks with fixed point,
      ylabel={Number of requests},
      xlabel={Claimed Chrome version},
      xticklabel style={font=\footnotesize},
      yticklabel style={font=\footnotesize},
      ylabel style={font=\small},
      xlabel style={font=\small},
      xmin=-2, xmax=132,
      ymin=1,
      ymax=50000,
      xtick pos=bottom,
      ytick pos=left,
      grid=major,
      grid style={dashed, gray!30},
      major grid style={line width=0.2pt},
      ymajorgrids=true,
      xmajorgrids=false,
    ]
    \addplot[
      fill=black,
      draw=black,
    ] coordinates {
      (0,7) (1,3) (2,10) (3,9) (4,87) (5,61) (6,29) (7,15) (8,10) (9,24) (10,35) (11,31) (12,34) (13,40) (14,42) (15,13) (16,22) (17,316) (18,13) (19,23) (20,11) (21,5) (22,11) (23,29) (24,158) (25,3) (26,11) (27,369) (28,125) (29,112) (30,33) (31,81) (32,106) (33,34) (34,78) (35,147) (36,186) (37,250) (38,5) (39,4328) (40,5281) (41,4731) (42,4819) (43,4692) (44,4586) (45,4804) (46,4630) (47,4771) (48,7095) (49,5690) (50,4586) (51,4802) (52,4611) (53,4687) (54,4719) (55,5266) (56,4938) (57,4814) (58,5381) (59,4384) (60,7040) (61,68) (62,162) (63,301) (64,66) (65,332) (66,763) (67,421) (68,467) (69,249) (70,337) (71,138) (72,285) (73,219) (74,724) (75,307) (76,279) (77,360) (78,2335) (79,751) (80,1761) (81,1741) (82,109) (83,410) (84,585) (85,498) (86,760) (87,1211) (88,3524) (89,1861) (90,737) (91,3206) (92,593) (93,344) (94,523) (95,1462) (96,794) (97,284) (98,676) (99,1570) (100,827) (101,1403) (102,346) (103,2269) (104,2425) (105,2063) (106,2190) (107,2483) (108,2959) (109,2475) (110,1984) (111,1977) (112,2011) (113,655) (114,2177) (115,38) (116,17114) (117,6628) (118,12296) (119,25793) (120,38068) (121,2398) (122,3077) (123,3308) (124,3981) (125,631) (126,2198) (127,497) (128,642) (129,1021) (130,1418)
    };
  \end{axis}
\end{tikzpicture}

%% file: parts/conclusion.tex
\section{Conclusion} \label{sec:conclusion}

In this paper, we presented a lightweight, passive approach to bot detection that combines user-agent coherence analysis with favicon-based heuristics. Unlike active techniques such as JavaScript challenges or CAPTCHAs, our method operates entirely on standard web server logs and requires no client-side interaction, preserving the user experience for legitimate visitors.

Our evaluation of over 4.6 million requests demonstrates that the proposed method detects 67.6\% of bot traffic while maintaining a false-positive rate of only 3.0\%. These results indicate that a large proportion of active crawlers deliberately spoof or omit identifying information in their user-agent strings and that checks are necessary to surface them.

For website operators seeking to detect or block bots, a layered strategy is recommended. As a first line of defense, our user-agent coherence method can be deployed directly in web server configurations or as lightweight middleware with minimal computational overhead. The favicon heuristic provides a complementary behavioural signal, particularly effective when combined with periodic cache invalidation. Requests that pass both filters can be considered likely human, while flagged requests can be routed to active challenges such as CAPTCHAs \cite{goos_captcha_2003} or proof-of-work systems \cite{back_hashcash_2002, noauthor_techarohqanubis_2026}, thereby restricting their friction to a smaller, genuinely ambiguous subset of traffic.

Looking ahead, the increasing volume of bot traffic-accounting for roughly half of all web requests-necessitates a shift in perspective. Rather than treating all bot access as adversarial, website operators could offer differentiated service tiers. Lightweight page versions, stripped of JavaScript, stylesheets, and visual assets, could be served to identified crawlers, substantially reducing server load and bandwidth costs while still providing the content they seek. Future work will focus on defining an API and a website representation to facilitate machine readability (for example, for LLM training).

%% file: parts/ethical_consideration.tex
\section{Ethical considerations} \label{sec:ethical-considerations}

This research strictly adheres to ethical guidelines and principles to ensure the responsible conduct of research. Due to the nature of our research, which involves analyzing web traffic and user-agent data, several measures were taken to protect the privacy and confidentiality of individuals and organizations involved in the study, including anonymizing data (using cryptography based privacy systems \cite{xu_prefix-preserving_2002}) and removing personally identifiable information before any kind of analysis.

%% file: parts/appendix.tex
\section{Appendix} \label{sec:appendix}

\begin{figure*}
  \centering
  \includegraphics[width=\textwidth]{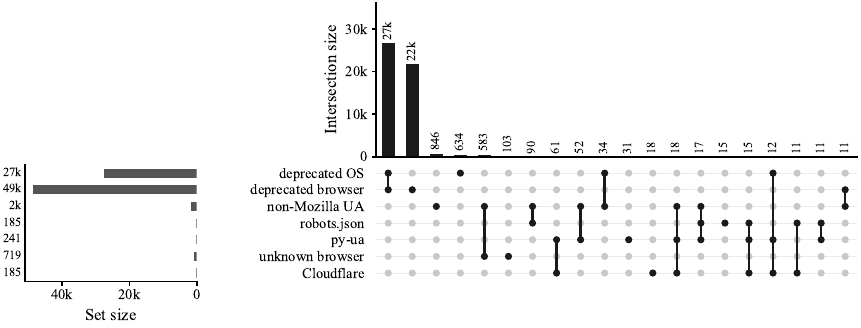}
  \caption{Overlap among bot detection methods across 54,945 unique user-agent strings, visualized with an UpSet plot \cite{lex_upset_2014}. The two largest intersections, deprecated OS $\cap$ deprecated browser (27k) and deprecated browser alone (22k), dominate, indicating that the majority of user-agent strings are caught using version-based heuristics.}
  \label{fig:unweighted-bots-upsetplot}
\end{figure*}

\begin{figure*}
  \centering
  \includegraphics[width=\textwidth]{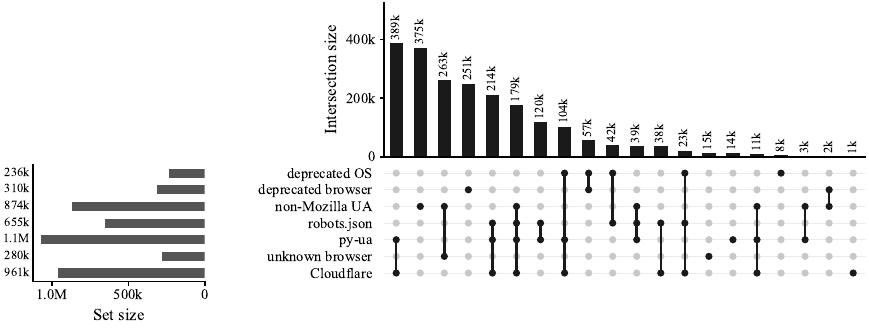}
  \caption{Overlap among bot detection methods weighted by request frequency. When weighted, the non-Mozilla UA and \texttt{py-ua} \cite{ong_selwinpython-user-agents_2026} methods gain prominence (874k and 1.1M set sizes, respectively), revealing that a smaller number of bot user-agent strings with non-standard prefixes or known bot signatures generate disproportionately high traffic volumes.}
  \label{fig:weighted-bots-upsetplot}
\end{figure*}

Figures \ref{fig:unweighted-bots-upsetplot} and \ref{fig:weighted-bots-upsetplot} present the distributions of detected user-agent string classifications, reported unweighted and weighted by the number of requests, respectively. Comparing the two figures reveals a notable asymmetry: while deprecated browser and OS versions account for the majority of distinct bot user-agent strings, a relatively small set of user-agents with non-Mozilla prefixes or known bot signatures generates disproportionately high traffic volumes. This suggests that the most aggressive bots rely on simple, poorly forged user-agents rather than diverse spoofing strategies.

%% file: refs.bib
@article{kwon2025web,
  title   = {Web-scraping AI bots disrupt databases and journals},
  author  = {Kwon, Diana},
  journal = {Nature},
  volume  = {642},
  pages   = {281},
  year    = {2025},
  doi     = {https://doi.org/10.1038/d41586-025-01661}
}

@article{Arlitt2001shopping,
  author     = {Arlitt, Martin and Krishnamurthy, Diwakar and Rolia, Jerry},
  title      = {Characterizing the scalability of a large web-based shopping system},
  year       = {2001},
  issue_date = {Aug. 2001},
  publisher  = {Association for Computing Machinery},
  address    = {New York, NY, USA},
  volume     = {1},
  number     = {1},
  issn       = {1533-5399},
  url        = {https://doi.org/10.1145/383034.383036},
  doi        = {10.1145/383034.383036},
  journal    = {ACM Trans. Internet Technol.},
  month      = aug,
  pages      = {44–69},
  numpages   = {26}
}

@inproceedings{liu_somesite_2025,
  location   = {New York, {NY}, {USA}},
  title      = {Somesite I Used To Crawl: Awareness, Agency and Efficacy in Protecting Content Creators From {AI} Crawlers},
  isbn       = {979-8-4007-1860-1},
  url        = {https://dl.acm.org/doi/10.1145/3730567.3732913},
  doi        = {10.1145/3730567.3732913},
  series     = {{IMC} '25},
  shorttitle = {Somesite I Used To Crawl},
  eventtitle = {{IMC} '25:{ACM} Internet Measurement Conference},
  pages      = {78--99},
  booktitle  = {Proceedings of the 2025 {ACM} Internet Measurement Conference},
  publisher  = {Association for Computing Machinery},
  author     = {Liu, Enze and Luo, Elisa and Shan, Shawn and Voelker, Geoffrey M. and Zhao, Ben Y. and Savage, Stefan},
  urldate    = {2026-02-11},
  date       = {2025-10-15}
}

@article{tiffany_maybe_2021,
  title    = {Maybe You Missed It, but the Internet 'Died' Five Years Ago},
  url      = {https://www.theatlantic.com/technology/archive/2021/08/dead-internet-theory-wrong-but-feels-true/619937/},
  journal  = {The Atlantic},
  author   = {Tiffany, Kaitlyn},
  urldate  = {2026-02-16},
  year     = {2021},
  month    = aug,
  day      = {31},
  date     = {2021-08-31},
  langid   = {english}
}

@online{belson_2025_2025,
  title      = {The 2025 Cloudflare Radar Year in Review- the rise of {AI}, post-quantum, and record-breaking {DDoS} attacks},
  url        = {https://blog.cloudflare.com/radar-2025-year-in-review/},
  titleaddon = {The Cloudflare Blog},
  author     = {Belson, David},
  urldate    = {2026-02-16},
  date       = {2025-12-15},
  langid     = {english}
}

@software{noauthor_techarohqanubis_2026,
  title     = {{TecharoHQ}/anubis},
  author    = {{Techaro}},
  rights    = {{MIT}},
  url       = {https://github.com/TecharoHQ/anubis},
  publisher = {Techaro},
  urldate   = {2026-02-16},
  date      = {2026-02-16},
  note      = {original-date: 2025-03-17T17:35:28Z}
}

@inproceedings{kim_scrapers_2025,
  location   = {New York, {NY}, {USA}},
  title      = {Scrapers Selectively Respect robots.txt Directives: Evidence From a Large-Scale Empirical Study},
  isbn       = {979-8-4007-1860-1},
  url        = {https://dl.acm.org/doi/10.1145/3730567.3764471},
  doi        = {10.1145/3730567.3764471},
  series     = {{IMC} '25},
  shorttitle = {Scrapers Selectively Respect robots.txt Directives},
  eventtitle = {{IMC} '25:{ACM} Internet Measurement Conference},
  pages      = {541--557},
  booktitle  = {Proceedings of the 2025 {ACM} Internet Measurement Conference},
  publisher  = {{ACM}},
  author     = {Kim, Taein and Bock, Karstan and Luo, Claire and Liswood, Amanda and Poroslay, Chloe and Wenger, Emily},
  urldate    = {2026-02-16},
  date       = {2025-10-28},
  langid     = {english}
}

@online{mueller_how_2025,
  title      = {How crawlers impact the operations of the Wikimedia projects},
  url        = {https://diff.wikimedia.org/2025/04/01/how-crawlers-impact-the-operations-of-the-wikimedia-projects/},
  titleaddon = {Diff},
  author     = {Mueller, Birgit and Danis, Chris and Lavagetto, Giuseppe},
  urldate    = {2026-02-17},
  date       = {2025-04-01},
  langid     = {american}
}

@software{ong_selwinpython-user-agents_2026,
  title    = {selwin/python-user-agents},
  rights   = {{MIT}},
  url      = {https://github.com/selwin/python-user-agents},
  author   = {Ong, Selwin},
  urldate  = {2026-02-23},
  date     = {2026-02-23},
  note     = {original-date: 2013-01-05T00:44:11Z}
}

@online{belinski_android_nodate,
  title    = {Android {API} Levels},
  url      = {https://apilevels.com/},
  author   = {Belinski, Eugene},
  urldate  = {2026-03-04},
  langid   = {english}
}

@software{noauthor_ai-robots-txtairobotstxt_2026,
  title     = {ai-robots-txt/ai.robots.txt},
  rights    = {{MIT}},
  url       = {https://github.com/ai-robots-txt/ai.robots.txt},
  publisher = {ai.robots.txt},
  urldate   = {2026-02-26},
  date      = {2026-02-26},
  note      = {original-date: 2024-03-27T17:48:29Z}
}

@online{noauthor_user-agent_2025,
  title      = {User-Agent header - {HTTP} {\textbar} {MDN}},
  url        = {https://developer.mozilla.org/en-US/docs/Web/HTTP/Reference/Headers/User-Agent},
  titleaddon = {{MDN} Web Docs},
  urldate    = {2026-03-04},
  date       = {2025-10-28},
  langid     = {american}
}

@online{noauthor_bots_nodate,
  title    = {Bots Directory {\textbar} Cloudflare Radar},
  url      = {https://radar.cloudflare.com/bots/directory},
  urldate  = {2026-03-05},
  langid   = {american}
}

@online{noauthor_verified_2025,
  title      = {Verified bots},
  url        = {https://developers.cloudflare.com/bots/concepts/bot/verified-bots/},
  titleaddon = {Cloudflare Docs},
  urldate    = {2026-03-05},
  date       = {2025-11-17},
  langid     = {english}
}

@online{noauthor_bots_2026,
  title      = {Bots},
  url        = {https://developers.cloudflare.com/bots/concepts/bot/},
  titleaddon = {Cloudflare Docs},
  urldate    = {2026-03-05},
  date       = {2026-02-05},
  langid     = {english}
}

@article{lex_upset_2014,
  title        = {{UpSet}: Visualization of Intersecting Sets},
  volume       = {20},
  rights       = {https://ieeexplore.ieee.org/Xplorehelp/downloads/license-information/{OAPA}.html},
  issn         = {1077-2626},
  url          = {http://ieeexplore.ieee.org/document/6876017/},
  doi          = {10.1109/TVCG.2014.2346248},
  shorttitle   = {{UpSet}},
  pages        = {1983--1992},
  number       = {12},
  journaltitle = {{IEEE} Transactions on Visualization and Computer Graphics},
  shortjournal = {{IEEE} Trans. Visual. Comput. Graphics},
  author       = {Lex, Alexander and Gehlenborg, Nils and Strobelt, Hendrik and Vuillemot, Romain and Pfister, Hanspeter},
  urldate      = {2026-03-09},
  date         = {2014-12-31}
}

@inproceedings{xu_prefix-preserving_2002,
  title      = {Prefix-preserving {IP} address anonymization: measurement-based security evaluation and a new cryptography-based scheme},
  issn       = {1092-1648},
  url        = {https://ieeexplore.ieee.org/document/1181415},
  doi        = {10.1109/ICNP.2002.1181415},
  shorttitle = {Prefix-preserving {IP} address anonymization},
  eventtitle = {10th {IEEE} International Conference on Network Protocols, 2002.},
  pages      = {280--289},
  booktitle  = {10th {IEEE} International Conference on Network Protocols, 2002. Proceedings.},
  author     = {Xu, Jun and Fan, Jinliang and Ammar, M.H. and Moon, S.B.},
  urldate    = {2026-03-11},
  date       = {2002-11},
  note       = {{ISSN}: 1092-1648}
}

@report{fielding_http_2022,
  title       = {{HTTP} Semantics},
  url         = {https://datatracker.ietf.org/doc/rfc9110},
  doi         = {10.17487/RFC9110},
  number      = {{RFC} 9110},
  institution = {Internet Engineering Task Force},
  type        = {Request for Comments},
  author      = {Fielding, Roy T. and Nottingham, Mark and Reschke, Julian},
  urldate     = {2026-03-11},
  date        = {2022-06},
  note        = {Num Pages: 194}
}

@report{nielsen_hypertext_1999,
  title       = {Hypertext Transfer Protocol – {HTTP}/1.1},
  url         = {https://datatracker.ietf.org/doc/rfc2616},
  doi         = {10.17487/RFC2616},
  number      = {{RFC} 2616},
  institution = {Internet Engineering Task Force},
  type        = {Request for Comments},
  author      = {Nielsen, Henrik and Mogul, Jeffrey and Masinter, Larry M. and Fielding, Roy T. and Gettys, Jim and Leach, Paul J. and Berners-Lee, Tim},
  urldate     = {2026-02-23},
  date        = {1999-06},
  note        = {Num Pages: 176}
}

@online{noauthor_user-agent-reduction_2025,
  title      = {User-Agent reduction - {HTTP} {\textbar} {MDN}},
  url        = {https://developer.mozilla.org/en-US/docs/Web/HTTP/Guides/User-agent_reduction},
  titleaddon = {{MDN} Web Docs},
  urldate    = {2026-03-11},
  date       = {2025-11-03},
  langid     = {american}
}

@inproceedings{senol_unveiling_2023,
  location   = {Copenhagen Denmark},
  title      = {Unveiling the Impact of User-Agent Reduction and Client Hints: A Measurement Study},
  isbn       = {979-8-4007-0235-8},
  url        = {https://dl.acm.org/doi/10.1145/3603216.3624965},
  doi        = {10.1145/3603216.3624965},
  shorttitle = {Unveiling the Impact of User-Agent Reduction and Client Hints},
  eventtitle = {{CCS} '23: {ACM} {SIGSAC} Conference on Computer and Communications Security},
  pages      = {91--106},
  booktitle  = {Proceedings of the 22nd Workshop on Privacy in the Electronic Society},
  publisher  = {{ACM}},
  author     = {Senol, Asuman and Acar, Gunes},
  urldate    = {2026-03-11},
  date       = {2023-11-26},
  langid     = {english}
}

@online{noauthor_user-agent_nodate,
  title   = {User-Agent Reduction},
  url     = {https://www.chromium.org/updates/ua-reduction/},
  urldate = {2026-03-11}
}

@software{noauthor_wicgua-client-hints_2026,
  title     = {{WICG}/ua-client-hints},
  url       = {https://github.com/WICG/ua-client-hints},
  publisher = {Web Incubator {CG}},
  urldate   = {2026-03-11},
  date      = {2026-03-01},
  note      = {original-date: 2018-10-24T12:20:52Z}
}

@online{noauthor_user-agent_nodate-1,
  title   = {User-Agent Client Hints},
  url     = {https://wicg.github.io/ua-client-hints/},
  urldate = {2026-03-11}
}

@online{noauthor_what_nodate,
  title      = {What Is Googlebot},
  url        = {https://developers.google.com/search/docs/crawling-indexing/googlebot},
  titleaddon = {Google for Developers},
  urldate    = {2026-03-13},
  langid     = {english}
}

@online{noauthor_overview_nodate,
  title   = {Overview of {OpenAI} Crawlers},
  url     = {https://developers.openai.com/api/docs/bots/},
  urldate = {2026-03-13},
  langid  = {english}
}

@online{andersen_history_2008,
  title      = {History of the browser user-agent string},
  url        = {https://webaim.org/blog/user-agent-string-history/},
  titleaddon = {{WebAIM}},
  author     = {Andersen, Aaron},
  urldate    = {2026-03-13},
  date       = {2008-09-03}
}

@software{stenberg_curlcurl_1996,
  title     = {curl/curl},
  rights    = {{MIT}},
  url       = {https://github.com/curl/curl},
  publisher = {curl},
  author    = {Stenberg, Daniel},
  urldate   = {2026-03-13},
  date      = {1996},
  note      = {original-date: 2010-03-18T22:32:22Z}
}

@online{stenberg_curl_1998,
  title      = {curl - Changes in 4.5.1},
  rights     = {{MIT}},
  url        = {https://curl.se/ch/4.5.1.html},
  titleaddon = {curl},
  type       = {documentation},
  author     = {Stenberg, Daniel},
  urldate    = {2026-03-13},
  date       = {1998-06-12}
}

@software{noauthor_nagios-pluginsnagios-plugins_2026,
  title     = {nagios-plugins/nagios-plugins},
  rights    = {{GPL}-3.0},
  url       = {https://github.com/nagios-plugins/nagios-plugins},
  publisher = {Nagios Plugins Project},
  urldate   = {2026-03-13},
  date      = {2026-03-11},
  note      = {original-date: 2014-01-16T20:40:26Z}
}

@software{noauthor_monitoring-pluginsmonitoring-plugins_2026,
  title     = {monitoring-plugins/monitoring-plugins},
  rights    = {{GPL}-3.0},
  url       = {https://github.com/monitoring-plugins/monitoring-plugins},
  publisher = {Monitoring Plugins},
  urldate   = {2026-03-13},
  date      = {2026-02-23},
  note      = {original-date: 2011-11-08T23:13:33Z}
}

@online{bamberg_favicon_2025,
  title      = {Favicon},
  url        = {https://developer.mozilla.org/en-US/docs/Glossary/Favicon},
  titleaddon = {{MDN} Web Docs},
  type       = {documentation},
  author     = {Bamberg, Will},
  urldate    = {2026-03-13},
  date       = {2025-07-11}
}

@online{google_stable_nodate,
  title      = {Stable Channel Update},
  url        = {https://chromereleases.googleblog.com/2014/11/stable-channel-update_18.html},
  date       = {2014-11-18},
  titleaddon = {Chrome Releases},
  author     = {{Google}},
  urldate    = {2026-03-18},
  langid     = {english}
}

@online{f5_web_nodate,
  title      = {Web Server {HTTP} Header Internal {IP} Disclosure},
  url        = {https://community.f5.com/discussions/technicalforum/web-server-http-header-internal-ip-disclosure/263036},
  author     = {{F5}},
  urldate    = {2026-03-18},
  langid     = {english}
}

@online{noauthor_http_nodate,
  title   = {{HTTP} {TRACE} method on {vCenter} port 9084},
  url     = {https://knowledge.broadcom.com/external/article/315560/http-trace-method-on-vcenter-port-9084.html},
  urldate = {2026-03-18}
}

@online{ascimoo_most_2018,
  title      = {The most important {HTTP} headers for scraping},
  url        = {https://go-colly.org/articles/scraping_related_http_headers/},
  author     = {{ascimoo}},
  urldate    = {2026-03-19},
  date       = {2018-01-01},
  langid     = {english}
}

@misc{wikimedia_foundation_user_2026,
  title     = {User Agent Breakdowns},
  url       = {https://analytics.wikimedia.org/dashboards/browsers/},
  publisher = {https://analytics.wikimedia.org/published/datasets/periodic/reports/metrics/browser/},
  author    = {{Wikimedia Foundation}},
  date      = {2026}
}

@online{tijhof_browser_2023,
  title      = {Browser adoption rates},
  url        = {https://timotijhof.net/posts/2023/browser-adoption/},
  titleaddon = {Timo Tijhof},
  type       = {Personal},
  author     = {Tijhof, Timo},
  urldate    = {2026-03-22},
  date       = {2023-02-16},
  langid     = {british}
}

@incollection{goos_captcha_2003,
  location    = {Berlin, Heidelberg},
  title       = {{CAPTCHA}: Using Hard {AI} Problems for Security},
  volume      = {2656},
  isbn        = {978-3-540-39200-2},
  url         = {http://link.springer.com/10.1007/3-540-39200-9_18},
  doi         = {10.1007/3-540-39200-9_18},
  shorttitle  = {{CAPTCHA}},
  pages       = {294--311},
  booktitle   = {Advances in Cryptology — {EUROCRYPT} 2003},
  publisher   = {Springer Berlin Heidelberg},
  author      = {Von Ahn, Luis and Blum, Manuel and Hopper, Nicholas J. and Langford, John},
  editor      = {Biham, Eli},
  editorb     = {Goos, Gerhard and Hartmanis, Juris and Van Leeuwen, Jan},
  editorbtype = {redactor},
  urldate     = {2026-03-23},
  date        = {2003},
  note        = {Series Title: Lecture Notes in Computer Science}
}

@online{shet_are_2024,
  title      = {Are you a robot? Introducing “No {CAPTCHA} {reCAPTCHA}”},
  url        = {https://security.googleblog.com/2014/12/are-you-robot-introducing-no-captcha.html},
  shorttitle = {Are you a robot?},
  titleaddon = {Google Online Security Blog},
  author     = {Shet, Vinay and {reCAPTCHA}},
  urldate    = {2026-03-23},
  date       = {2024-12-03},
  langid     = {english}
}

@tech{back_hashcash_2002,
  title    = {Hashcash - A Denial of Service Counter-Measure},
  author   = {Back, Adam},
  date     = {2002-08-01},
  langid   = {english}
}

@article{husak_https_2016,
  title        = {{HTTPS} traffic analysis and client identification using passive {SSL}/{TLS} fingerprinting},
  volume       = {2016},
  issn         = {1687-417X},
  url          = {https://doi.org/10.1186/s13635-016-0030-7},
  doi          = {10.1186/s13635-016-0030-7},
  pages        = {6},
  number       = {1},
  journaltitle = {{EURASIP} Journal on Information Security},
  shortjournal = {{EURASIP} J. on Info. Security},
  author       = {Husák, Martin and Čermák, Milan and Jirsík, Tomáš and Čeleda, Pavel},
  urldate      = {2026-03-25},
  date         = {2016-02-26},
  langid       = {english}
}

@inproceedings{wei_deep_2019,
  location  = {Cham},
  title     = {A Deep Learning Approach to Web Bot Detection Using Mouse Behavioral Biometrics},
  isbn      = {978-3-030-31456-9},
  doi       = {10.1007/978-3-030-31456-9_43},
  pages     = {388--395},
  booktitle = {Biometric Recognition},
  publisher = {Springer International Publishing},
  author    = {Wei, Ang and Zhao, Yuxuan and Cai, Zhongmin},
  editor    = {Sun, Zhenan and He, Ran and Feng, Jianjiang and Shan, Shiguang and Guo, Zhenhua},
  date      = {2019},
  langid    = {english}
}

@online{noauthor_bot_nodate,
  title      = {Bot Manager},
  url        = {https://www.akamai.com/products/bot-manager},
  titleaddon = {Akamai},
  urldate    = {2026-03-25},
  langid     = {english}
}

@online{noauthor_cloudflare_nodate,
  title    = {Cloudflare Bot Management Solution},
  url      = {https://www.cloudflare.com/en-gb/application-services/products/bot-management/},
  urldate  = {2026-03-25},
  langid   = {english}
}

@online{edwards_open_2025,
  title      = {Open source devs say {AI} crawlers dominate traffic, forcing blocks on entire countries},
  url        = {https://arstechnica.com/ai/2025/03/devs-say-ai-crawlers-dominate-traffic-forcing-blocks-on-entire-countries/},
  titleaddon = {Ars Technica},
  author     = {Edwards, Benj},
  urldate    = {2026-03-25},
  date       = {2025-03-25},
  langid     = {english}
}

@software{noauthor_arcjetwell-known-bots_2026,
  title     = {arcjet/well-known-bots},
  rights    = {{MIT}},
  url       = {https://github.com/arcjet/well-known-bots},
  publisher = {Arcjet},
  urldate   = {2026-03-25},
  date      = {2026-03-18},
  note      = {original-date: 2024-09-06T13:46:25Z}
}

@online{lepage_new_2017,
	title = {New in Chrome 61},
	url = {https://developer.chrome.com/blog/new-in-chrome-61},
	titleaddon = {Chrome for Developers},
	type = {Blog},
	author = {Lepage, Pete},
	urldate = {2026-03-27},
	date = {2017-09-05},
	langid = {english},
}

@software{matomo_matomo-orgdevice-detector_2026,
	title = {matomo-org/device-detector},
	rights = {{LGPL}-3.0},
	url = {https://github.com/matomo-org/device-detector},
	publisher = {Matomo Analytics},
	author = {{Matomo}},
	urldate = {2026-03-27},
	date = {2026-03-27},
	note = {original-date: 2014-04-02T21:41:32Z},
}

@software{beech_jaybizzlecrawler-detect_2026,
	title = {{JayBizzle}/Crawler-Detect},
	rights = {{MIT}},
	url = {https://github.com/JayBizzle/Crawler-Detect},
	author = {Beech, Mark},
	urldate = {2026-03-27},
	date = {2026-03-26},
	note = {original-date: 2015-03-23T20:05:37Z},
}

@software{rohm_anthemakergoodbots_2026,
	title = {{AnTheMaker}/{GoodBots}},
	url = {https://github.com/AnTheMaker/GoodBots},
	author = {Röhm, Anton},
	urldate = {2026-03-27},
	date = {2026-03-27},
	note = {original-date: 2023-01-03T18:04:23Z},
}

@online{mueller_quo_2026,
	title = {Quo Vadis, Crawlers? Progress and what’s next on safeguarding our infrastructure},
	url = {https://diff.wikimedia.org/2026/03/26/quo-vadis-crawlers-progress-and-whats-next-on-safeguarding-our-infrastructure/},
	shorttitle = {Quo Vadis, Crawlers?},
	titleaddon = {Diff},
	author = {Mueller, Birgit and Danis, Chris and Lavagetto, Giuseppe},
	urldate = {2026-03-28},
	date = {2026-03-26},
	langid = {american},
}

@report{koster_robots_2022,
	title = {Robots Exclusion Protocol},
	url = {https://datatracker.ietf.org/doc/rfc9309},
	doi = {10.17487/RFC9309},
	number = {{RFC} 9309},
	institution = {Internet Engineering Task Force},
	type = {Request for Comments},
	author = {Koster, Martijn and Illyes, Gary and Zeller, Henner and Sassman, Lizzi},
	urldate = {2026-03-28},
	date = {2022-09},
	note = {Num Pages: 12},
}
